# A Method to Measure the Flatness of the LSST Focal Plane Assembly in Situ[1]

Willy Langeveld

*Stanford Linear Accelerator Center*
*Stanford University*
*Stanford, CA*

**ABSTRACT**

In this note I describe an inexpensive and simple laser-based method to measure the flatness of the LSST focal plane assembly (FPA) in situ, i.e. while the FPA is inside its cryostat, at $-100^0$ C and under vacuum. The method may also allow measurement of the distance of the FPA to lens L3, and may be sensitive enough to measure gravity- and pressure-induced deformations of L3 as well. The accuracy of the method shows promise to be better than 1 micron.

---

[1] Work supported in part by the Department of Energy contract DE-AC02-76SF00515

# 1. Introduction

The focal plane assembly (FPA) of the Large Synoptic Survey Telescope (LSST) [1] consists of an integrating structure, on which are mounted 25 rafts which hold 9 CCD sensors each, except for the corner rafts which have only 3 sensors each (see figure 1). The total number of sensors is thus 201. During normal operation the FPA is located inside a cryostat and kept at $-100^0$ C (to limit sensor dark current) and under vacuum (for insulation purposes). The flatness tolerance of the FPA is 10 microns (peak-to-valley). The problem addressed here is how to measure compliance when the device is mounted in the telescope and ready for operation, i.e. in situ. A complication arises because the FPA is located only 25 mm from the inside of lens L3 [1].

In this note, we assume that there already exists an ex situ method to measure the flatness of the array of sensors before and/or after it has been mounted in its cryostat, e.g. using a scanning laser triangulation apparatus. "Ex situ" in this sense means that the measurement takes place prior to complete assembly of the rest of the camera and attachment to the telescope. This then serves to generate a baseline measurement of the flatness of the array.

This note describes an in situ, inexpensive and simple laser-based method to measure changes in the flatness of the FPA with respect to the baseline measurement using the FPA itself.

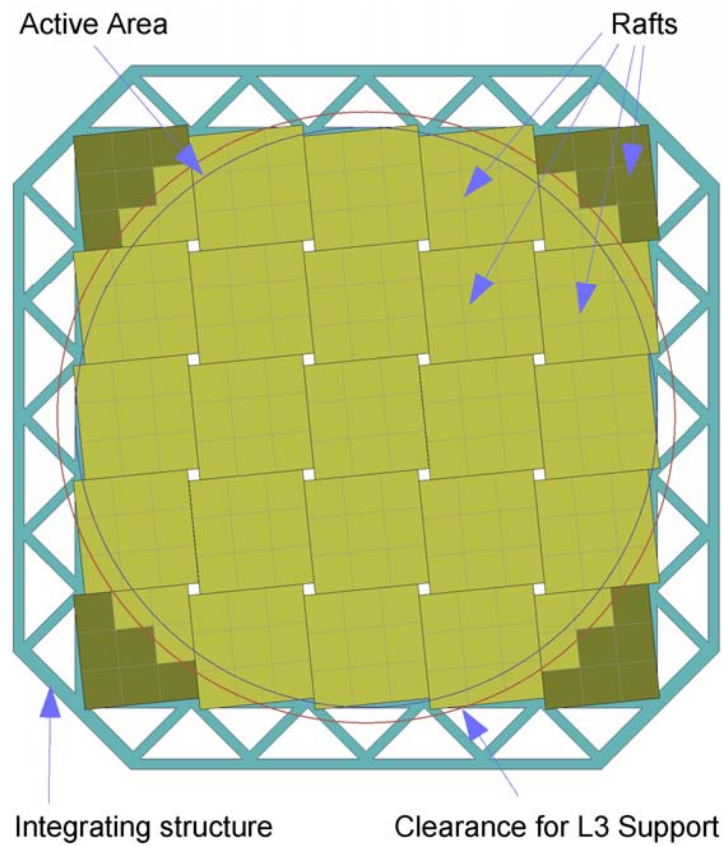

Figure 1  Integrating structure with 25 rafts of sensors.

## 2. The Basic Method

To measure the changes in flatness in situ, one or more laser assemblies together with a suitable set of optical elements and diffractive pattern generators is mounted on the integrating structure, such that the secondary beams emerging from the diffraction gratings can project onto the FPA a large set of spots. After the baseline measurement, the lasers and the FPA sensors are turned on and the FPA is read out. The location of each spot is measured using a standard centroiding algorithm and these nominal positions are stored in a database. Whenever desired, the lasers can be turned on again, and the positions can be compared to the nominal positions stored in the database. From the differences, one can derive the deviations of the focal plane flatness from the baseline using a suitable software algorithm.

Moreover, it can be arranged that some secondary laser beams hit lens L3 and some of the light reflects back onto the FPA. One can find out (by careful measurement) which spots on the FPA are due to reflections off L3 (see figure 2). In principle it is thus possible to measure the distance of the FPA to the bottom of L3. If a sufficient number of such reflected spots can be measured, it may be possible to detect distortions of L3 due to gravity (as the telescope moves to different azimuths) and due to the air pressure outside the cryostat.

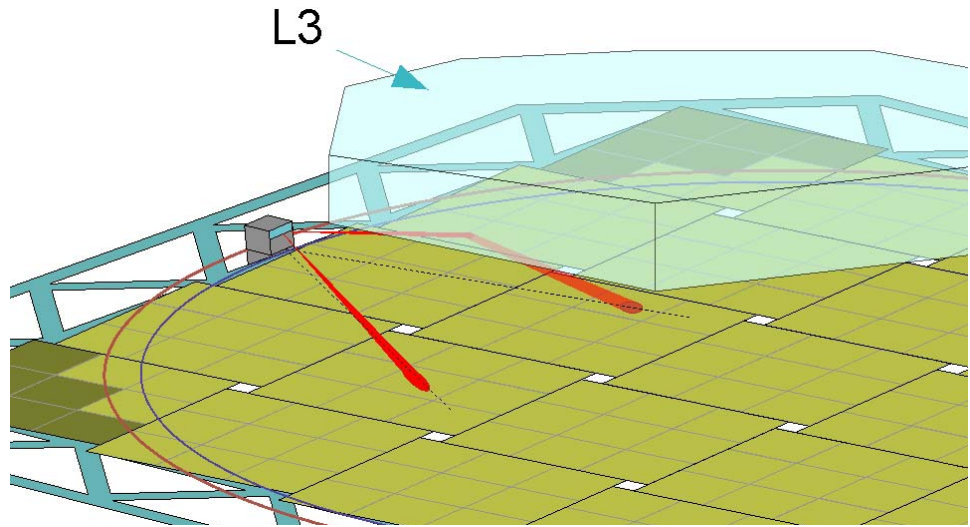

Figure 2  Directly produced spot and a bounce off L3. The spot shapes are different.

## 3. Spot Measurement and Accuracy

The centroid of a spot in an image can be determined very accurately. Even with rather poor signal to noise ratio, the accuracy can be better than $1/30^{th}$ of a pixel size, and accuracies of $1/1000^{th}$ of a pixel have been achieved in astronomical context [2]. With 10 micron pixel size, that translates to an accuracy of, at least, better than 0.3 microns. Since the secondary laser beams intersect the FPA at a shallow angle, the sensitivity to vertical excursions is increased by an additional factor. The method described here should therefore, in principle, be sufficiently accurate and accuracy will probably be limited by other factors.

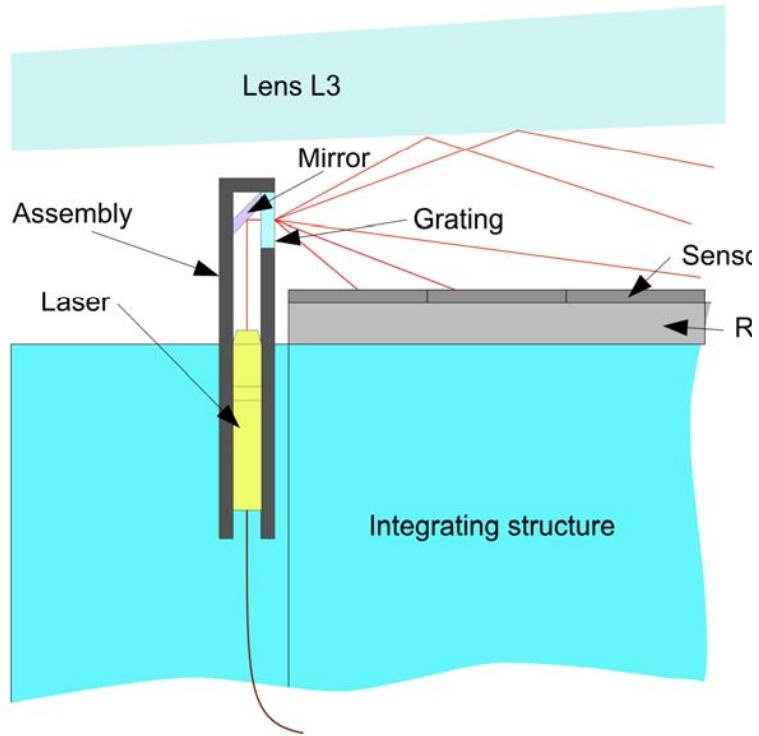

Figure 3  Possible laser assembly.

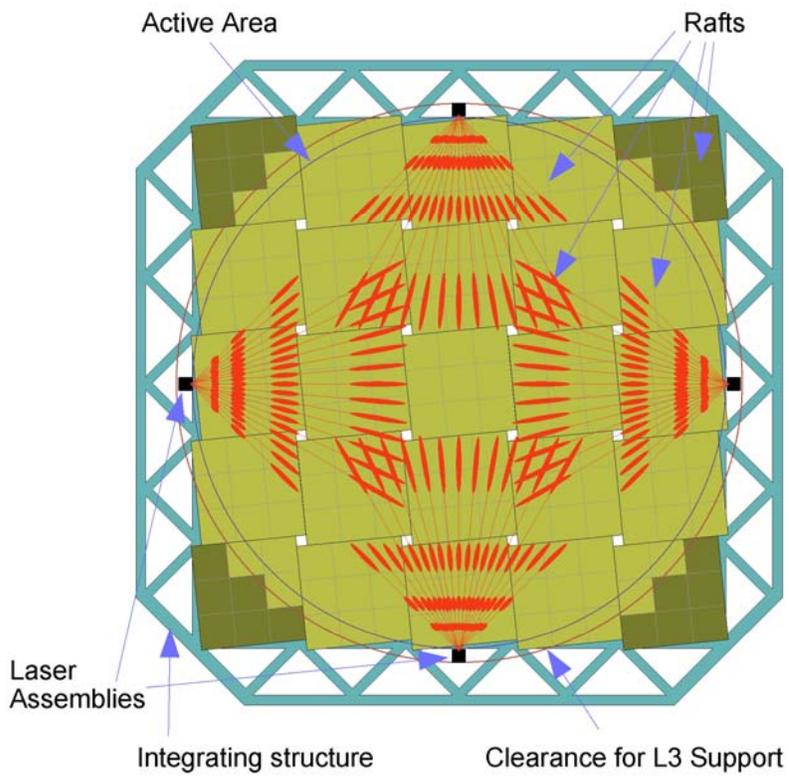

Figure 4  Integrating structure with possible locations for the laser assemblies.

Since the spots are formed from secondary beams incident at a shallow angle, the spot shape is roughly that of an ellipse. Measuring the lengths of the minor and major axis allows one to calculate the angle of incidence, and hence one can distinguish spots that originate directly from the grating from those that are due to a bounce of the secondary beam off lens L3. It is possible that secondary beams partially reflect off the surface of the FPA, then off L3 and back onto the FPA. Such multiple reflections can be distinguished by their decreasing intensity and reconstructed angle of incidence.

One can also measure the orientation of the ellipse: the major axis will point back to the grating from which the corresponding secondary beam originated.

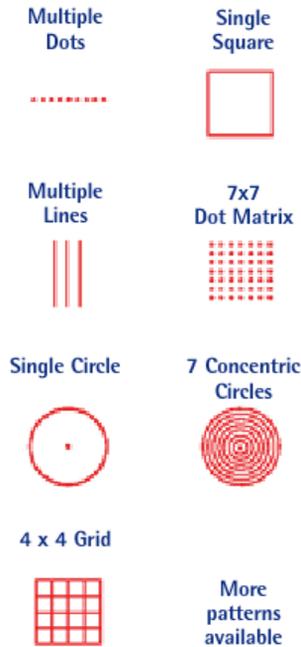

Figure 5  Available diffractive pattern generators.

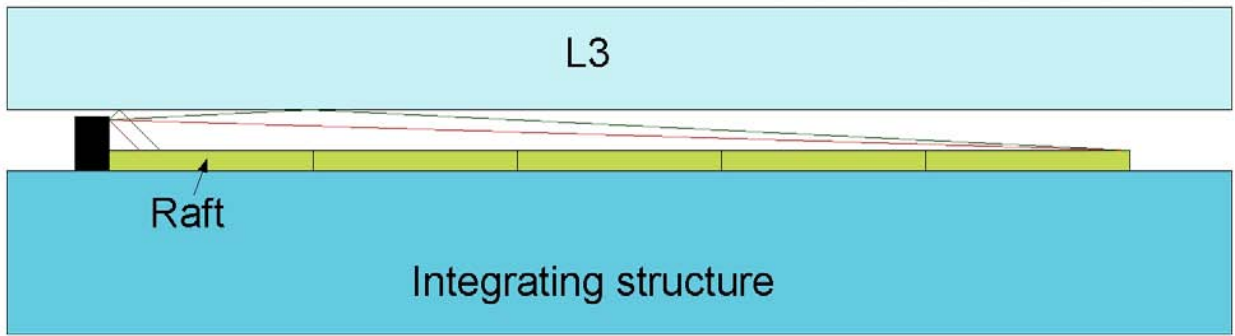

Figure 6  To-scale diagram showing which secondary beams will intersect the FPA at angles shallower than 45 degrees.

## 4. Laser Assemblies and Gratings

One possible way to arrive at a grid of spots is using a set of four or more laser assemblies. A possible laser assembly is shown in figure 3. A relatively standard laser (e.g. a red or green diode laser) shines onto a suitable pattern-generating diffraction grating followed by a diagonal mirror.

A possible arrangement of the laser assemblies on the integrating structure can be seen in figure 4. The red circle (diameter 680 mm [3]) indicates the inner diameter of the support ring for L3. In the current design, there is only about 20 mm space between the edge of the support ring and the active area (indicated by the blue circle, diameter 640 mm [1]).

Diffractive pattern generators are available from a number of sources. For example, Edmund Scientific has several standard patterns available, such as those shown in figure 5. Custom gratings can be fabricated to generate any desired pattern [4].

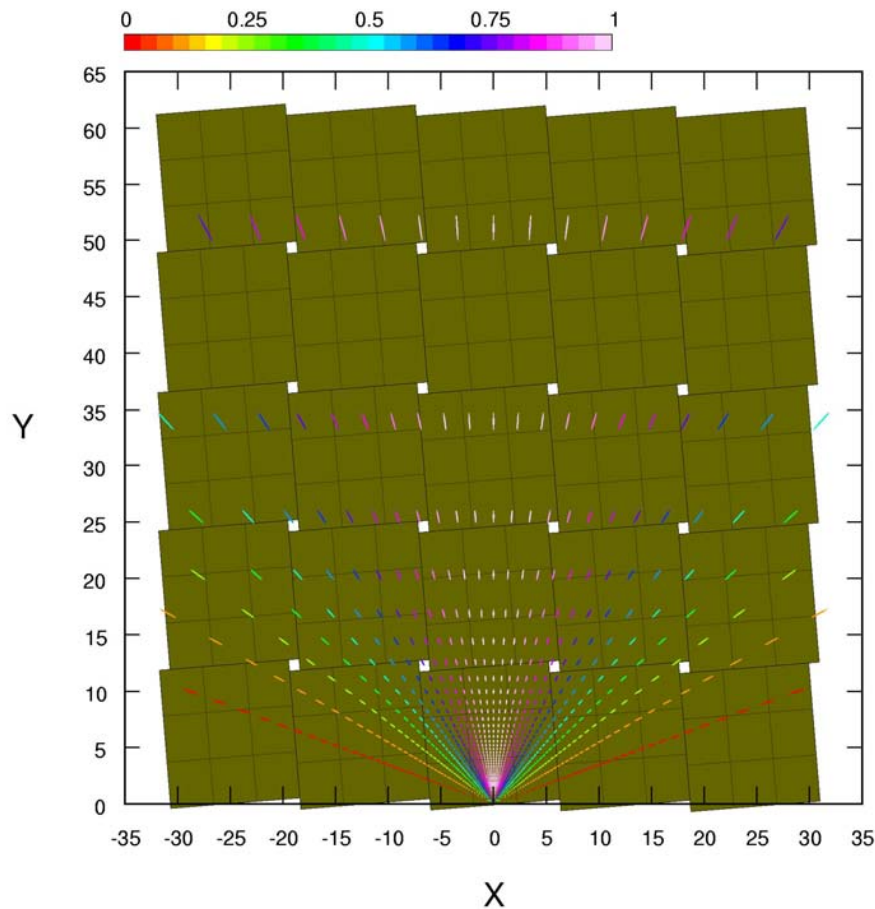

Figure 7  Simulated diffraction pattern of grating with 50 apertures/mm vertically and 150 apertures/mm horizontally. The scale shows the normalized intensity.

From the grating equation [5]:

$$d \sin \theta = p \lambda$$

with $d$ the distance between apertures, $p$ an integer and $\lambda$ the wavelength of the laser light, we can derive that the angle between the $0^{th}$ and first principal maxima is:

$$\theta_{01} \approx \lambda / d$$

for $d \gg \lambda$. Let us assume that the laser beam diameter is about 1 mm and that the origin of the secondary beams lies 23 mm above the FPA surface (see figure 6). In order to be able to separate spots near the grating that have 45 degrees or shallower angle of incidence, we then need roughly one principal maximum per degree of angle for the longitudinal spot distribution on the FPA (i.e. away from the grating). This leads to a grating with $d \approx 50 \lambda$. If $\lambda \approx 0.45$ microns, that means $d \approx 20$ microns. In the $y$ direction the grating should therefore have roughly 50 apertures per mm. The lateral spot distribution would need to be sparser in order to be able to separate spots close to the grating. One spot per three degrees satisfies the criterion. In the $x$ direction the grating should therefore have about 150 apertures per mm.

The intensity distribution for the spots resulting from a grating with aperture width $a$ and aperture spacing $d$ is given by [5]:

$$I = \frac{1}{N^2} \left( \frac{\sin \alpha}{\alpha} \right)^2 \left( \frac{\sin N\beta}{\sin \beta} \right)^2$$

where

$$\alpha = \frac{\pi a \sin \theta}{\lambda}, \qquad \beta = \frac{\pi d \sin \theta}{\lambda}.$$

In order to have a reasonably equal intensity for all spots, $a$ should be small, which occurs when the aperture width $a$ is small compared to $d$, i.e. when grating is close to ideal.

The results of a simulation of the spot pattern resulting from a two-dimensional diffraction grating with 50 apertures/mm vertically and 150 aperture/mm horizontally, with $a = 0.1$ microns is shown in figure 7. The colors of the spots indicate the (normalized) intensity.

## 5. Using Custom Optical Elements

A possible criticism to the system outlined above is that it is difficult to guess what happens to a laser assembly such as depicted in figure 3 when it is cooled to $-100^0$ C. We will come back to this in the next section. Meanwhile, we consider a variation on the basic theme by using a single laser and four custom optical elements. These optical elements could be made from essentially single pieces of a low-expansion-coefficient material such as fused silica, with (perhaps integral) diffraction gratings, directly mounted onto the integrating structure. To ensure equal brightness of the four spot patterns, the first element would have a 25% reflecting diagonal, the second 33%, the third 50% and the last 100%. A not-to-scale schematic of the arrangement is shown in figure 8.

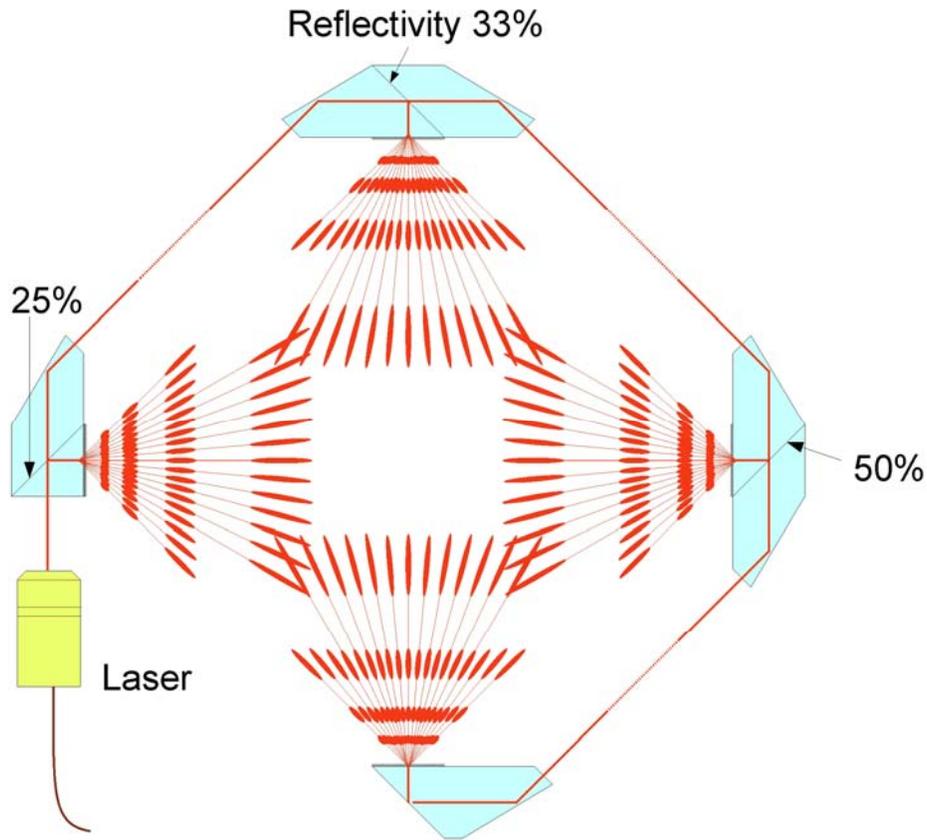

Figure 8  Alternative setup using one laser and four custom optical elements.

## 6. Laser Stability

A weak point in both scenarios is the stability of the laser itself. There are in fact several issues to be considered: will a laser even work at $-100^0$ C? Military specifications for lasers in e.g. laser range finders require operability to $-32^0$ C, and storage temperatures down to $-57^0$ C [6]. It is therefore not clear that lasers work at the LSST operating temperature. Also, since the laser will need to be operated in a vacuum (I will consider the possibility of an external laser in a moment), there is a question about adequate cooling. One might think that the laser will only need to be operated for a few seconds, but one would probably want to give the laser some time to warm up and find a stable equilibrium. Therefore, the laser might need to have a cooling strap attached to its housing, and perhaps a thermometer to measure when the temperature has stabilized. It is likely that the time needed to warm up the laser to its operating temperature is longer than the time needed to illuminate the FPA, and therefore the laser might need to be equipped with a shutter, so that it does not illuminate the FPA while warming up. Alternatively, a heater could be provided which keeps the laser at a stable temperature. Both of these options would raise a question as to whether the laser assembly could realistically be mounted on the integrating structure: it would not be prudent to heat up the integrating structure locally.

All these issues lead one to consider the possibility of using an external laser, i.e. a laser that shines through a suitable glass porthole into the cryostat. The laser could be mounted on the side of the cryostat and shine in horizontally from the left onto the first optical element in figure 8, or it could be mounted on the rear of the cryostat and follow a somewhat more complicated path to the first optical element. It might even be located such that it shines through lens L3 onto a suitable mirror arrangement. The advantages for the laser are clear: it is always used at room temperature and at atmospheric pressure, it is somewhat more accessible, it does not need a cooling strap or a heater and there is no need for more cables penetrating the cryostat wall. The disadvantage is also clear: the laser would not be firmly attached to the integrating structure of the FPA, and therefore it would be difficult if not impossible to repeat the spot position measurement to within a few microns, especially considering the fact that the FPA will be actuated to compensate for atmospheric turbulence. Whether the actuation system can reposition the FPA to a nominal position to within a micron or two is a current item of further study [7].

So the question becomes: is it really necessary for the laser beam itself to be precisely aligned from one measurement to the next? While precise laser alignment would be very convenient, I think that it is not necessary: small horizontal and/or vertical displacements and small angle variations of the laser beam lead to very specific distortions of the spot pattern, and can probably be accounted for in software.

It therefore seems that the driving factor for an accurate measurement of the spot locations is the intrinsic stability of the laser during a single measurement, rather than precise repeatability of the laser beam position and angle from one measurement to the next. This then argues for an external laser, since operating conditions can then be controlled more easily. One would run the laser continuously (or turn it on long enough for it to stabilize) and use a shutter for the exposure.

It is of course possible to use four (or more) external lasers, each of which shines through a porthole (or L3) in the cryostat directly (or through a set of mirrors) onto a diffraction grating. This makes the pattern produced by each laser independent of the others. Whether that is an advantage or a disadvantage is not entirely clear. A computer simulation could be performed to see which is best.

## 7. Advantages and Disadvantages

Overall, a system such as the ones described above has many advantages. It can be inexpensive: the total cost of lasers, assemblies, mirrors and gratings should be less than a few thousand dollars, even assuming custom gratings. Using completely custom optical elements might be more expensive, but perhaps not prohibitively so. It is easy to implement: there is room on the integrating structure, and no fancy machining is needed. The rafts are completely unaffected by this system, since the lasers and/or optical elements are mounted directly on the integrating structure (or in the case of external lasers, on the cryostat). The system uses the front of the FPA sensors to measure the flatness of the array directly. The system has the ability to measure the distance of the front of the FPA to L3, and may even be able to measure distortions in L3. For the internal laser option, only a small number of cables need to be routed and fed through the cryostat wall. For any external laser a porthole needs to be provided (or an arrangement of mirrors such that it can shine through L3), but no cable feed-throughs.

The one overall disadvantage of the system is that it requires the FPA sensors to be turned on and read out. In normal operating mode, CCDs of the type considered for the LSST cannot be read out while at room temperature. The dark current is so high that the CCDs saturate. The dark current starts to become manageable when the temperature drops below -20 to -40$^0$ C [7]. Problems with the stability of a laser or laser assembly when cooled to low temperatures and operated under vacuum conditions may exist, but can perhaps be worked around and in any case lead to predictable changes in the spot pattern. It may be easier to employ external lasers, at the cost of less exact repeatability of the measurements, but it should be possible to account for the differences in software.

This is to be contrasted with some of the other systems currently under discussion. One system ("optical straight edge") requires the installation of four CCD cameras and four beam splitters on each of the 25 rafts for a total of 100 cameras and beam splitters, as well as 10 lasers. These 100 CCD cameras will all need to be read out and special software will need to be written to handle them. The FPA flatness is measured only indirectly and the measurement is insensitive to changes internal to the raft and the sensors mounted thereupon. Design of this system is complicated by the fact that its laser beams need to pass between rafts uninterrupted, and features need to be designed into the rafts to allow for affixing and precisely positioning the CCD cameras and beam splitters. Cables from the 100 cameras will need to be routed and fed through the cryostat wall. Stability of the lasers under the environmental conditions is also a concern here.

Another system currently being studied ("laser interferometry") requires mechanical fixtures for attaching 75 optical fibers to the integrating structure and feeding them through the cryostat wall. Also, three reflectors will need to be mounted on the bottom of each raft, a total of 75. This system, as well, only measures the flatness of the FPA indirectly. The system would employ an external laser.

It is, of course, also possible to have both the system proposed here and one of the alternative systems at the same time.

## 8. Summary

In this note I have described a conceptual system of one or more lasers and/or optical elements to project onto the LSST FPA a large set of elliptical spots. The (centroid) locations, orientations and aspect ratios of these ellipses can provide detailed information about the flatness of the FPA when compared to a baseline ex-situ measurement. Spots formed after a bounce of secondary beams off L3 can be used to measure the distance of L3 to the FPA, and perhaps will allow measurement of deformations of L3.

Areas for further study are:
1. Can a run-of-the-mill laser be operated at $-100^0$ C?
2. What is the optimal diffraction grating?
3. What is the required power level for the laser, and what is the best wavelength?
4. Does the laser need a remote-control shutter or is the laser stable enough that it can just be turned on and off?
5. Do spots from secondary beam bounces off L3 have sufficient intensity?
6. Can an external laser be mounted with sufficient stability?
7. What is the cost of custom-made optical elements?
8. What is the best material to use for such optical elements?
9. What is the cost of a porthole in the side or rear of the cryostat?
10. Can external lasers shine through L3?
11. Should one use one laser and multiple custom optical elements or multiple lasers and simpler optical elements?
12. What software is needed to analyze the spot data and arrive at vertical deviations of the FPA and distances of the FPA from L3?

I wish to thank Eric Lee, Kirk Gilmore and Layton Hale for helpful discussions.